# Name order and the top elite: Long-term effects of a hidden curriculum

Eiji YAMAMURA[1*]

1. *Department of Economics, Seinan Gakuin University, 6-2-92 Sawaraku Nishijin Fukuoka 814-8511, Japan.*

* Corresponding author

E-mail: yamaei@seinan-gu.ac.jp

1. Conceptualization, Formal analysis, Investigation, Methodology, Data curation, review, Writing – original draft preparation.

**Funding:** No.

**Conflicts of interest/Competing interests:** There is no conflict of interest.

# Abstract

In Japanese primary and secondary schools, an alphabetical name list is used in various situations. Generally, students are called on by the teacher during class and in the ceremony if their family name is early on the list. Therefore, students whose surname initials are earlier in the Japanese alphabet acquire more experience. Surname advantages are considered to have a long-term positive effect on life in adulthood. This study examines the surname effect.

The data set is constructed by gathering lists of representative figures from various fields. Based on the list, we calculate the proportion of surname groups according to Japanese alphabetical column lists. The major findings are as follows: (1) people whose surnames are in the “A” column (the first column among 10 Japanese name columns) are 20% more likely to appear among the ruling elite but are less likely to appear in entertainment and sports lists. (2) This tendency is rarely observed in the University of Tokyo entrance examination pass rate. Consequently, the “A” column surnames are advantageous in helping students succeed as part of the elite after graduating from universities but not when gaining entry into universities. The surname helps form non-cognitive skills that help students become part of the ruling elite instead of specific cognitive skills that help students enter elite universities.

# 1. Introduction

Surnames are related to life outcomes in various fields. A surname being combined with inherited physical aptitude leads to specialties in certain sports activities [1,2]. As reflected in the distribution of surnames among the elite and underclass, social status is more strongly inherited than physical aptitude [3]. Apart from heredity, learning performance in school would be an outcome of the advantage of surnames. The surnames' alphabetical order was correlated with academic performance in high school [4]. However, even before entering high school, non-cognitive skills formed in childhood play a critical role in improving quality of life outcomes [5–7]. During schooling, people obtained cognitive skills through the official (academic) curriculum and non-cognitive skills through the "hidden (non-academic) curriculum." The "hidden curriculum" has long-term effects later in life through worldview and preference formation [8,9].

As a kind of "hidden curriculum," an alphabetical surname list is subconsciously or unintentionally used by teachers in various situations in Japan [10–14]. For instance, students who appeared earlier on the list are called on by teachers during class and in ceremonies first. Consequently, students whose surnames appear earlier in the Japanese alphabet gained more experience during life at school. In contrast, people whose surnames are at the end of the list are inclined to be left behind and feel discriminated against [15]. Alphabetical ordering's name effect disincentivizes teamwork, which has a detrimental effect on output [16].

Various studies investigated the influence of the surname in various settings namely, firm performance[17–19], peer nomination[20], election[21–25], and citations in the academic world [26–28]. Essentially, a surname's earlier alphabetical order is positively related to better performance and evaluation. The point is that people choose one at the top of the list to save time considering performance quality, academic papers, and the candidate's policy. These findings concerning the name order effects are similar to the finding that a paper's cover page is more likely to be cited [29–31]. This holds for schools as teachers tend to call on pupils at the top of the name list first.

Different professions involve different levels of social status and required ability. Therefore, the influence of name order on performance would differ between professions. So far, no studies have attempted to examine whether name order effects vary according to the type of profession. This study's purpose is to compare the effect of name order on ruling elites with a focus on entertainers and sports players. In this way, we attempted to analyze whether a surname during a school career has long-term effects. We collected lists of representative figures from elites, such as the highest position among members of the career elite in government offices, professors of the most prestigious university, leaders of academic society, and members of the Supreme Court, etc. In addition, the name list of those who passed the entrance examination of the most prestigious university, which mainly supplied Japanese society's elite. Hence, we can explore the timing of the name order effect observed. Apart from the elite, we also gather lists of the most eminent artists in the entertainment world and the most eminent players in the world of sports and professional board games.

The dataset allows us to analyze the name order effect by illustrating figures and conducting linear regression. We emphasized surnames in the first column ("A" column).

# 2 Data and Methods

## 2.1. Data collection

In Japanese alphabetical order, the "A" column is listed first although its order is different from the English alphabet. Within the "A" column, five Japanese characters ("Kana") are ordered as "A", "I", "U", "E", and "O." Next to the "A" column, "Ka" column comes and "Ka", "Ki", "Ku", "Ke," and finally "Ko." In total, there are 10 columns. The surnames are listed in the order of the 50-character "Kana" syllabary. Naturally, students whose surnames are found in the "A" column gained more abundant experience during their school careers.

We gathered the top 1500 surnames that cover 90% of the general population in Japan. The proportion of the name columns is illustrated in Fig. 1. The rate of the "A" column is around 22%, which is the highest among the ten columns (Fig.1). Next to "A," that of the "Ka" column is about 14%. As a whole, the rate of the columns decreases as their order becomes later.

**Fig 1. The ratio of the general population according to alphabetical columns.** This is the figure legend.

We collected lists of representative figures from various fields to construct the dataset. Due to the Act for the Protection of Personal Information, the lists are limited to information that is open to the public from various internet sources. The fields are roughly classified into two groups: the ruling and academic elite, and entertainment and sports celebrities.

In comparison to elites, entertainment and sports celebrities are less likely to hold privileged and powerful societal positions. In this way, the effect of surname order on ruling power is compared with that of non-elites. The elite group consists of 29 sub-cate-gories: academic elite (nobel prize laureates, members of the Japanese Academy, participants in the International Mathematical Olympiad), professors from the University of Tokyo, which is acknowledged as the most prestigious university (six departments, including Economics, Law, Literature, Medical, Science, and Engineering), directors of academic society (five societies, including Economics, Public law, Private law, Physics, Engineering), Supreme Court Justices, past vice ministers who are the top of bureaucrats (eight government agencies, including the Finance, Foreign, Education, Agriculture, Post, Health, Economics, Culture departments).

We aim to check at which point in time the surname effect on performance emerges by considering the distribution of surnames among elite students. Students of the University of Tokyo are generally expected to join the academic and ruling elite. Hence, as an elite group, we also include those who passed the entrance examination of the University of Tokyo (six departments, including Economics, Law, Literature, Science and Engineering, Agriculture and Medical). The total number of successful applicants is around 3000 every year. Successful applicants for the exam are mostly around 18 or 19 years of age and are poised to join the elite but are not part of the real elite yet.

There are 16 subcategories of entertainment and sports. The “Japan Record Academy Award” is presented to artists of the most popular song every year. In the ceremony, a Singer, Lyricist, and Composer received the award. Concerning the award, we can make 3 lists of past winners of the singer, lyrist, and composer awards. Similarly, the “Japan Film Academy Award” is presented to the most evaluated actor and film. Meanwhile, the 1980s was when comedy became popular in Japan. Many popular comedians appeared in “Oretachi Hyokinzoku (We are comical tribe),” the most popular TV comedy show from the 1980s. Based on the show, we made a list of comedians. For literature, there are two lists. We made two past winners lists of the Akutagawa Prize, which was presented to the most promising novice novelist of pure literature, and of the Naoki Prize, which was presented to the most promising novice novelist of popular literature [32]. As for board games, two lists are professional Japanese Chess (Shogi) and Igo. Turning to sports, there are 6 lists for Olympians from the 2020 Tokyo Olympics, a list of professional boxers of past world champions, professional baseball players, Sumo wrestlers in the highest class, and professional motorboat racers. They are representative figures in the field of popular culture and sports but do not have the ruling power.

Traditionally, in Japan women's surnames tended to change after getting married [33]. Therefore, a woman's surname in adulthood is likely to differ from her surname in childhood if she is married. Unfortunately, we cannot obtain marital status from the lists collected in this study. Furthermore, male students traditionally appeared earlier on lists than female ones in Japanese schools. That is, females were listed alphabetically despite being later than males. Take an example, Aoki is at the top in "A column." Nevertheless, a female having Aoki as her surname appears later than a male having the surname Watanabe which is the last "Wa" column. Gender discrimination has gradually changed after the 1980s [10–14]. However, naturally, there is a difference in name order effects between genders especially those who experienced school life before the 1990s.

Hence, in most of the lists, we exclude women from the list because we aim to examine the surname effect in school. The exceptions are lists of teenagers such as math-Olympian and successful applicants for the entrance exam. This is not only because her name was unlikely to change, but also because the data for math Olympians is limited after the 2000s and the outcome of the exam was from the results of 1997. Further, in the field of entertainment and literature, stage names or pen names are widely used. Therefore, we use their real surname to make the list, rather than a well-known stage name or a pen name.

The ability is a necessity to join the elite in 29 lists and is considered to differ from that which is required to join the ranks of entertainers and sports players. A high educational background is necessary to join the elite. However, a high educational background is not required to be an entertainer and sports player. Using the data, we can investigate which fields the name order has a positive effect on.

## 2.2 Ethical issues

We collected data from various open sources. The data possess or pose no ethical problems such as health issues. As for sources where we collected data, see S1 Table in Appendix.

## 2.3 Method

Our model assesses how the surname is related to the outcome of social status. The simple estimated function takes the following form:

$Y_{ih} = \alpha_0 + \alpha_1$ *"A"column* $_{ih} + X_i B + u_i$

where $Y_{ih}$ is the outcome variable for work $i$ and name column $h$. α denotes the regression parameters. $u_{ih}$ is the error term. The estimation method is the Ordinary Least Square (OLS) model. In the basic specification, $Y_{it}$ is

$\widehat{x_{ih}} = (x_{ih} - \bar{x}_h),$

where $x_{ih}$ is the rate of surname column $h$ for work $I$ while $\bar{x}$ is the rate of surname column $h$ for the general population. The value $\widehat{x_{ih}}$ is positive when the work $i$'s rate in surname column *h* is higher than the rate in surname column *h* in the general population. The key variable is the *"A" column*. It is 1 if the column is "A," otherwise 0. The *"A" column* is positive if people with a surname in column "A" succeed in their chosen profession. Dummies for categories of work are included as X to control for the vector of categories. For instance, as explained in the previous section, work *i* contains 6 department Professors of the University of Tokyo. To control the group-specific factors, a dummy of the professors is included. Similarly, we include a dummy for the "Academic elite dummy" covering three groups, a dummy for "Directors of academic society", a dummy for Vice-ministers, a dummy for passing the entrance exam, a dummy for athletes, a dummy for literature authors, and a dummy for artists.

As shown in Fig. 1, there is a clear difference in rates between surname columns. Hence, we calculate its standardized values as follows:

$$\widetilde{x_{\imath h}} = (x_{ih} - \bar{x}_h)/\bar{x}_h,$$

In addition to $\widehat{x_{\imath h}}$, as an alternative specification, we also use $\widetilde{x_{\imath h}}$ to control the difference in the rate of surnames in the general population. The statistical software used in this study was Stata/MP 15.0.

## 2.4 Results

Fig. 2 demonstrates how the distribution of surnames using this study's sample differs from that of the general population. A positive (Negative) value implies that the ratio of the surname in this study's sample is larger (smaller) than that of the general population. The ratio of the "A" column is distinctly larger than 0. The value of the "A" column is about 2, meaning that people in the “A” column are larger by 2 % than the general population. Apart from the “A” column, only the “Na” column shows statistical differences and negative values.

Fig. 3 illustrates the difference in surname distribution from the general sample after splitting the sample into the Elite and entertainment-sports groups. Concerning the “A” column, we observe a striking difference in distribution between the two samples. Based on the subsample of the elite, the ratio of the “A” column is distinctly larger than 0 and its value is approximately 4. Therefore, the elite’s surname in the “A” column is larger by 4% than the general population. In contrast, in the sample of the entertainment-sports group, the surname ratio is lower than 0 despite being statistically insignificant. Besides the “A” column, there is no significant difference in the surname column between groups.

**Fig 2. Difference of ratio from the general population by Alphabetical Columns.** This is the figure legend.

**Fig 3. Comparison between ruling elite and entertainment-sports groups: Difference of ratio from the general population by Alphabetical Columns. This is the figure legend.**

Tables 1-3 show the results of regression estimation. In these Tables, the dependent variable is the percentage difference from the general population in columns (1) and (2), while being a standardized value of difference from the general population in columns (3) and (4). Further, in columns (1) and (2), the reference of the *“A” column* is the rest of the columns. Hence, the coefficient of the *“A” column* shows how the rate of the *“A” column* differs from other columns. The expected positive sign of the *“A” column* means

that rate of surnames in the *“A” column* is higher than in other columns. For closer examination, in columns (2) and (4), we set the “Ka” column as the reference group that is second in order and so appears next to the “A” column in the Japanese Alphabetical list. Then, in addition to a dummy for the *“A” column,* we also include dummies for other columns in confounder variables. Therefore, the expected positive sign of the *“A” column* means that rate of the surname in the *“A” column* is higher than in the “Ka” column.

Table 1 reports the results using a full sample covering the elite group and the entertainment-sports group. The *“A” column* shows a positive sign in all columns. However, statistical significance is observed only in column (1).

1 Table 1. Regression estimation (OLS model). Full sample

| | (1) EXCESS | (2) EXCESS | (3) Standardized EXCESS | (4) Standardized EXCESS |
|---|---|---|---|---|
| *“A” column* | 2.090*** (0.800) | 1.381 (1.091) | 0.040 (0.055) | 0.051 (0.06) |
| *“Ka” column* | | <Reference> | | <Reference> |
| *“Sa” column* | | − 0.996 (0.941) | | − 0.070 (0.067) |
| *“Ta” column* | | − 0.567 (0.938) | | − 0.040 (0.073) |
| *“Na” column* | | − 1.542* (0.878) | | −0.165** (0.075) |
| *“Ha” column* | | − 1.393 (0.992) | | − 0.115 (0.078) |
| *“Ma” column* | | − 0.363 (0.892) | | − 0.021 (0.069) |
| *“Ya” column* | | − 1.242 (0.907) | | − 0.128 (0.081) |
| *“Ra” column* | | − 0.224 (0.776) | | 0.368 (0.275) |
| *“Wa” column* | | − 0.053 (0.817) | | 0.274 (0.224) |
| R-square | 0.03 | 0.05 | 0.0002 | 0.04 |
| Observations | 450 | 450 | 450 | 450 |

2 Note: Numbers within parentheses are robust-standard errors.

3 EXCESS RATE is (Rate of the group’s population − Rate of the general population)

*p<0.10, **p<0.5, ***p<0.01

Table 2 reports the results using a subsample of the elite group. The *"A" column* shows a positive sign while being statistically significant in all columns. Therefore, the rate of surnames in the *"A" column* is higher than in other columns in the elite group. The absolute value of the coefficient of the *"A" column* is 0.206 in column (3). This implies that the person with a surname in the "A" column is 20.6% higher than the rate in the general population, compared with the rate of a person with a surname in other columns. Turning to the comparison between the "A" and "Ka" columns. The absolute value of the coefficient of the *"A" column* is 0.169 in column (4). This suggests that the person with a surname in the "A" column is 16.9% higher than the rate in the "K" column. Apart from the *"A" column,* the coefficient's signs are mixed. Further, statistical significance is not observed except for the *"NA" column* and *"HA" column.* That is, persons with surnames in the "Ka" columns are not more likely to appear in the list than others. The name order effect is limited to the top of the alphabetical list.

Table 2. Regression estimation (OLS model). Ruling elite sample

| | (1) EXCESS | (2) EXCESS | | (3) Standardized EXCESS | (4) Standardized EXCESS |
|---|---|---|---|---|---|
| *"A" column* | 4.306*** (0.932) | 3.773*** (1.278) | | 0.206*** (0.054) | 0.169** (0.074) |
| *"Ka" column* | | <Reference> | | | <Reference> |
| *"Sa" column* | | — 0.667 (1.176) | | | — 0.050 (0.084) |
| *"Ta" column* | | 0.519 (1.081) | | | 0.050 (0.083) |
| *"Na" column* | | — 1.470 (1.085) | | | —0.173* (0.097) |
| *"Ha" column* | | — 2.177* | | | — 0.193** |

| | | | | |
|---|---|---|---|---|
| | | (1.138) | | (0.088) |
| *“Ma” column* | | − 0.293<br>(1.061) | | − 0.025<br>(0.083) |
| *“Ya” column* | | − 0.658<br>(1.103) | | − 0.076<br>(0.102) |
| *“Ra” column* | | − 0.085<br>(0.887) | | 0.038<br>(0.076) |
| *“Wa” column* | | 0.039<br>(0.960) | | 0.100<br>(0.264) |
| R-square | 0.11 | 0.15 | 0.01 | 0.03 |
| Observations | 290 | 290 | 290 | 290 |

Note: Numbers within parentheses are robust-standard errors. Standardized EXCESS RATE is (Rate of group’s population − Rate of the general population)/ Rate of the general population.
*p<0.10, **p<0.5, ***p<0.01

Table 3 exhibits the results using a subsample of the entertainment-sports group. In all results, the *“A” column* shows a negative sign, while indicating statistical significance except for column (4). Interestingly, this is a remarkable contrast to the results of Table 2. The rate of the surname in the *“A” column* is lower than in other columns in the entertainment-sports group. The absolute value of the coefficient of the *“A” column* is 0.260 in column (3). Accordingly, a person with a surname in the *“A”* column is 26% lower than the rate in the general population. However, as shown in column (4), there is no difference between the *“A”* and *“Ka”* columns.

Table 3. Regression estimation (OLS model). Entertainment-sports sample

| | (1)<br>EXCESS | (2)<br>EXCESS | (3)<br>Standardized EXCESS | (4)<br>Standardized EXCESS |
|---|---|---|---|---|
| *“A” column* | −1.927**<br>(0.959) | −2.954*<br>(1.692) | −0.260**<br>(0.108) | −0.161<br>(0.105) |
| *“Ka” column* | | <Reference> | | <Reference> |
| *“Sa” column* | | − 1.575<br>(1.607) | | − 0.110<br>(0.112) |
| *“Ta” column* | | − 2.537<br>(1.081) | | − 0.206<br>(0.136) |
| *“Na” column* | | − 1.670 | | −0.140 |

| | | | | | |
|---|---|---|---|---|---|
| | | (1.515) | | | (0.117) |
| *"Ha" column* | | 0.027<br>(1.796) | | | 0.027<br>(0.139) |
| *"Ma" column* | | － 0.492<br>(1.637) | | | － 0.013<br>(0.126) |
| *"Ya" column* | | － 2.301<br>(1.514) | | | － 0.222<br>(0.138) |
| *"Ra" column* | | － 0.474<br>(1.495) | | | 0.967<br>(0.749) |
| *"Wa" column* | | － 0.221<br>(1.520) | | | 0.587<br>(0.409) |
| R-square | 0.03 | 0.08 | | 0.005 | 0.11 |
| Observations | 160 | 160 | | 160 | 160 |

Note: Numbers within parentheses are robust-standard errors. Standardized EXCESS RATE is (Rate of group's population－Rate of the general population)/ Rate of the general population.
*p<0.10, **p<0.5

Figs. 4 (a) and (b) illustrate the difference in the percentage of the "A" column between the specific groups and the general population. Figs 4 (a) and (b) demonstrate elite and entertainment-sports groups, respectively. Concerning the elite, positive values are observed for 26 among 29 lists. In most elite worlds, the rate of persons with a surname in the "A" column is higher than the general population. Interestingly, as shown in the bottom part of Fig 4(a), the pass rates for the University of Tokyo entrance exam are almost 0 regardless of department. Different from other groups in Fig 4(a), passing the entrance exam is the outcome at ages 18-20.

As opposed to the elite, as demonstrated in Fig 4(b), negative values are observed for 23 of the 16 lists. However, most of the cases show around 1-2 %, which is nearly 0 although some outliers such as the awarded singer indicate about －13 %.

In estimations, we put focus on the "A" column. As for the distribution of other columns in each work, see more detailed S2 Figures in S2 Appendix.

**Fig 4(a). Difference from the general population: Specific ruling elite groups. This is the figure legend.**

**Fig 4(b). Difference from the general population: Specific entertainment-sports groups. This is the figure legend.**

# 3 Discussion

## 3.1 Implication

This study examines how name order influenced life outcomes. In Japan, persons with surnames appearing first on school lists are likely to have various experiences because teachers call their names first in various situations. The surname in the first column in the Japanese alphabetical list has a distinct effect on life outcomes. We observe a significant gap compared with surnames in the lists' second column. This is consistent with the studies that found the cover page effect in academic citation [29–31].

The mechanism through which surnames influence life outcomes is not directly examined. As illustrated in Figure 4(a), professors in the most prestigious university tend to have a surname at the top of the alphabetical order. This outcome appears to be reflected by the many citations of their studies likely caused by name order on alphabetical lists [26–28]. However, this interpretation cannot hold for a successful career as a government officer. This supports the postulation that the surname advantage in school has a long-term effect on life outcomes. Another possibility is that the surname effect is inherited [1–3]. This inheritance seems to be positively correlated with the probability of attending a prestigious university. However, such a tendency is not observed.

The surname in the first column increases the probability of joining the ruling elite while reducing the probability of joining the ranks of popular entertainers and top athletes. Abundant experiences in school are thought to form cognitive and non-cognitive skills. More skilled persons are more able to display high performance and so succeed in various fields. However, this study's finding implies that skilled people tend to devote their labor toward joining the ruling elite, rather than becoming entertainers or athletes. They enjoy sports and cultural hobbies but do not become professionals in these fields even if they have enough skills to succeed. Instead, they achieve dominant positions in academia, legal circles, and as government officers.

Entering the University of Tokyo is the first step to becoming part of the ruling Japanese elite. Surprisingly, however, we hardly observed the name order effect. In our interpretation, the skills obtained through school experiences are not related to becoming a celebrity. The University of Tokyo's exam is a written test. The test only evaluates certain aspects such as a student's accurate knowledge or whether the student makes accurate calculations within a certain time. Meanwhile, the test does not evaluate personality, creativity, etc. Furthermore, passing the exam is the outcome around 18-19 years of age. The skill formed through experiences are useful in various situations in the elite's professional lives and contributed to success after graduating from university. This is consistent with previous studies providing evidence that the effect of childhood education on non-cognitive skills persisted even in adulthood, whereas the effect on cognitive skills disappeared [5–7].

Owing to the convenience of teachers calling students, an alphabetical list is used in Japan. Consequently, people's lives are unintentionally influenced by surname order on alphabetical lists. The advantage of surname order occurred naturally by coincidence. The unintended discrimination was criticized for gender inequality from the feminist viewpoint because male students' names appeared earlier than female ones in Japanese school lists [10–14]. However, the name order problem can be considered more generally because inequality caused by name order is also observed among males. Apart from the network of surname affinity [34], surname ordering possibly caused socio-economic segregation in society. To reduce inequality between people at the top of the list and others, calling students should be randomized to equalize opportunities of experiences in school life.

## 3.2 Strength

Previous studies examined the name order effect by considering specific situations and fields such as sports activity [1,2], performance in school [4], academic performance, academic world [26–28], firm performance [17–19], peer nomination [20], election [21–25]. Our study is the first to examine the surname effect using data covering various professions and fields and then compared the effect between different professions and fields. Then, different from previous studies, the surname effect differs between the ruling elite and entertainment-sports celebrities. This is the strength of using open data sources.

Previous studies found that surname is closely related to ethnicity [35,36]. According to the Japanese Population Census, foreign residents occupied only 2.2 % of the whole Japanese population in 2022 [37]. The advantage of the Japanese case is a more racially homogenous society than Western societies. The surname effect is hardly influenced by a diversity of races and ethnicities.

## 3.3 Limitation

We collected data from various open sources. However, the sample is small because we restricted the number of professions to construct the data. Besides this study’s sample, there are possibly various professions in which the name order effect can be examined. We admit that the sample selection is, to a certain extent, arbitrary. We need to extend the data to cover other professions to explore whether the finding is generally observed.

We can add other universities when we considered the proportion of professors' surnames. However, as is widely acknowledged, the University of Tokyo is the most prestigious in Japan. Therefore, we can justify limiting our study to the University of Tokyo to consider the surname effect among the top elite. This holds for other professions in the way of collecting samples.

Female surnames are far more likely to change after marriage in Japan. So, a married female's surname tends to differ from her childhood surname. Due to the limitation of data open to the public, we could not identify who was married. Further, we cannot obtain information about females' childhood surnames if they were married. The sample is mostly limited to males and so we cannot consider the surname effect on females. However, the traditional system that males were listed earlier than females has increasingly changed since the 1990s. The change of system arguably influenced the name order effects. However, we cannot consider the effect. Naturally, the estimation results possibly suffered from selection biases.

Due to open data limitations, we cannot conduct a detailed examination. Using individual-level data containing males and females, we should consider the surname effect in future research.

# 4 Conclusion

We compare the effect of the top names on the list by comparing outcomes for the ruling elite with the successful persons in the entertainment and sports fields. The list of the top surnames on the list leads students directly to positions as government officers and academicians, but not artists or sports players. Persons with surnames at the top of alphabetical lists have abundant experiences during their school tenure. The advantage the surname offered resulted in success in adulthood, rather than during their school career.

The advantage allows people to attain a higher social status. Then, the next generation can inherit this status [3]. Naturally, societal stratification could become entrenched. To mitigate social segregation and inequality, students should be called randomly to ensure that there is no reliance on alphabetical lists even if inconvenient for teachers. Furthermore, whether our findings apply to other countries needs to be verified.

# Supporting information

S1 Appendix. Sources where we gathered the dataset.

S2 Appendix. The supplementary materials provided the distribution of surname columns separately for 41 sub-categories of work.

# Acknowledgments

Iwould like to thank Editage (http://www.editage.com) for this manuscript’s English language editing and review.

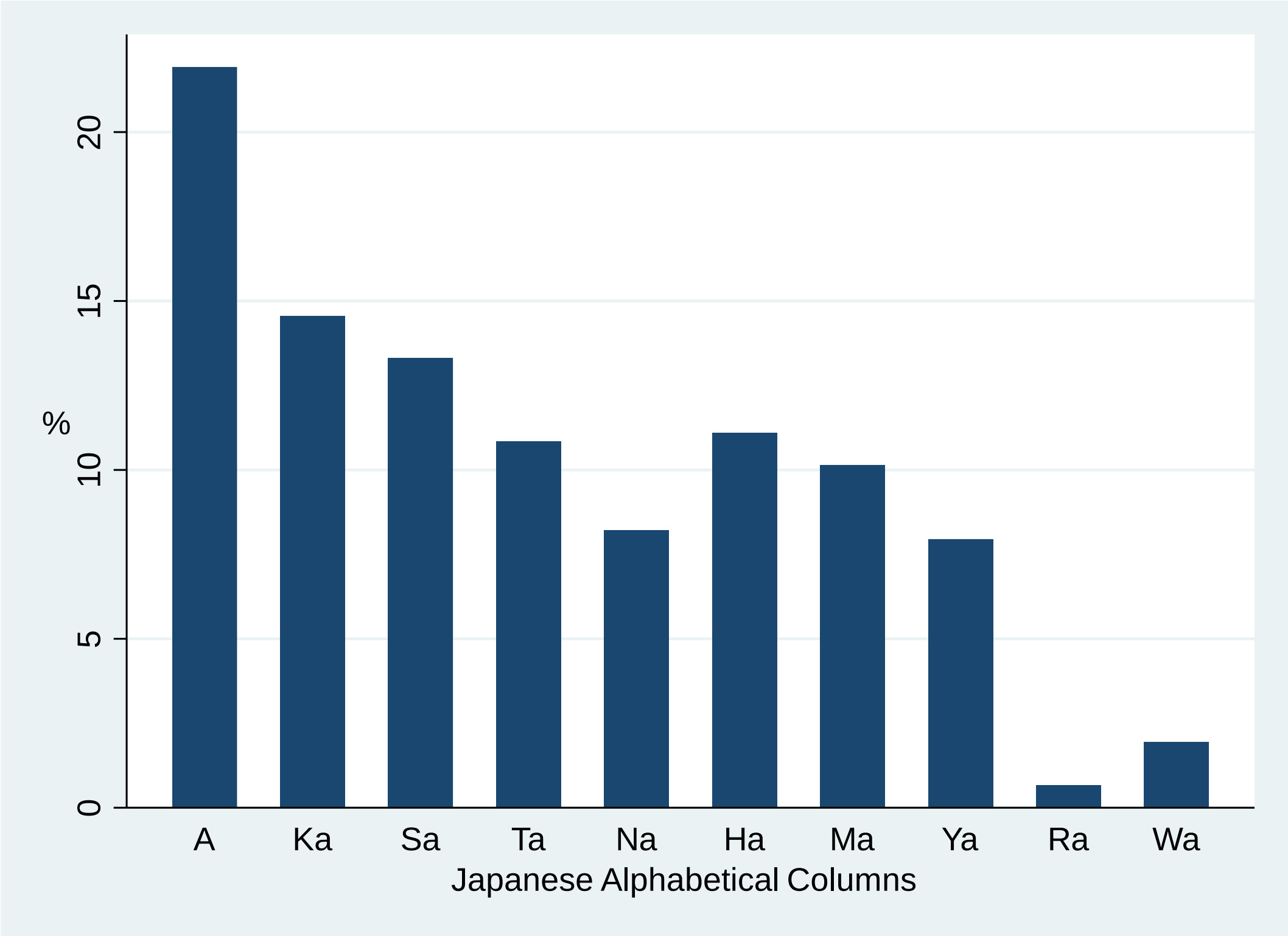

Fig 1. The ratio of the general population according to alphabetical columns.
Note: Data of the top 1500 family names are used to calculate percentages. This covers 9,3200,000 people which is about 75% of the whole Japanese population (12,570,000).
Sources: Family Name Ranking
https://myoji-yurai.net/prefectureRanking.htm (accessed on December 18, 2022).

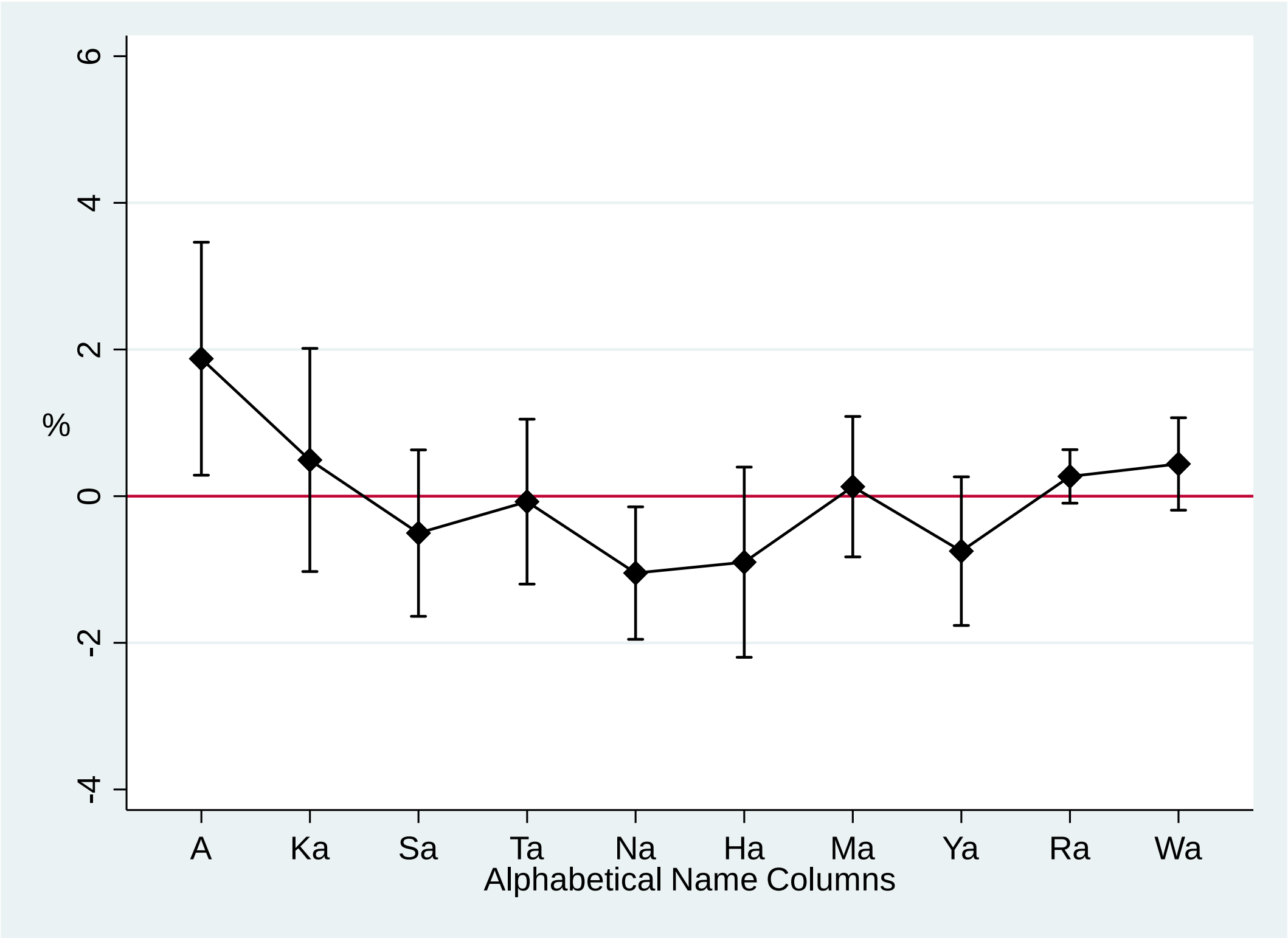

Fig.2. Difference of ratio from the general population by Alphabetical Columns
Note: Vertical line (0) is the reference group (General population)

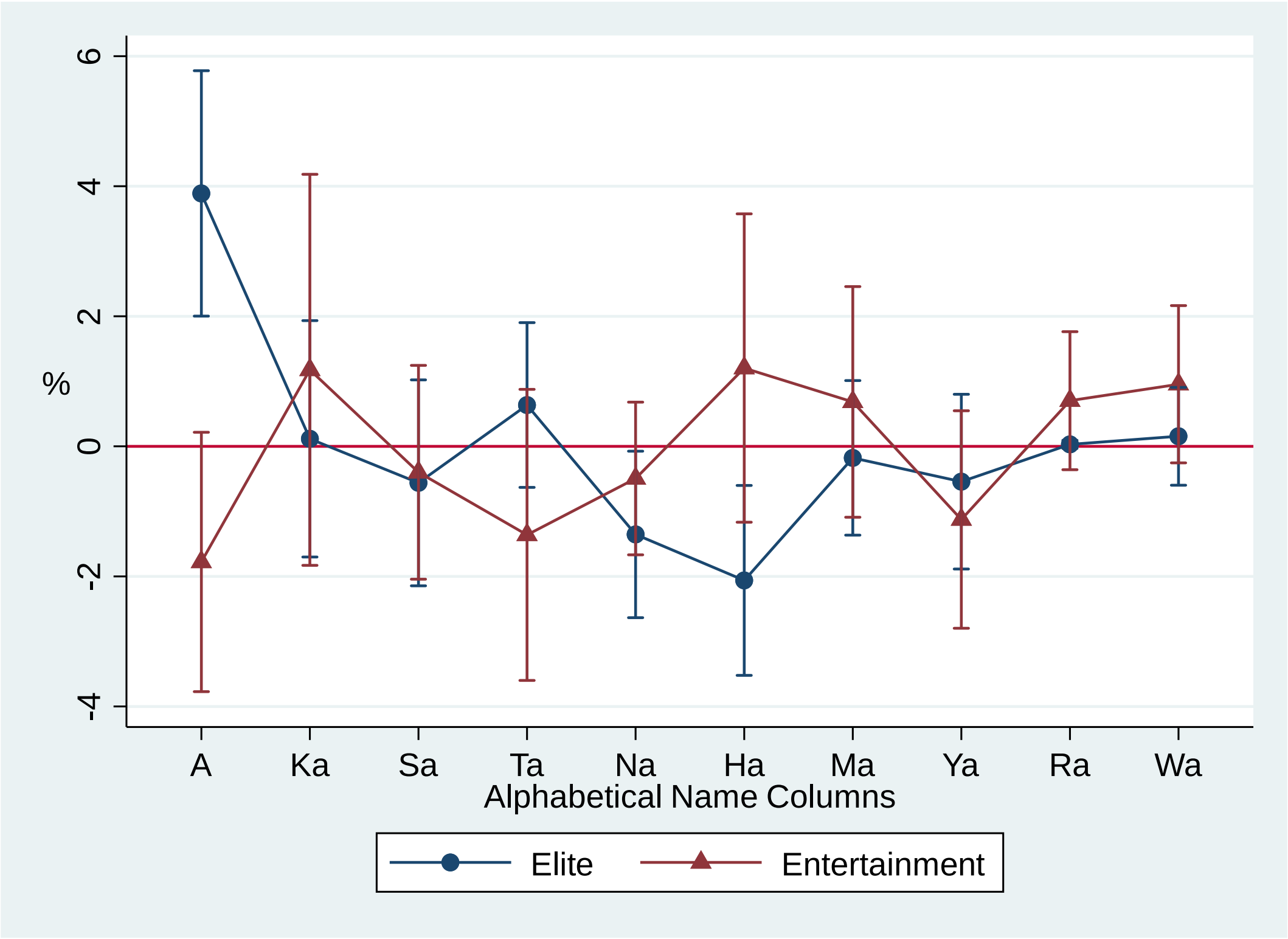

Fig.3. Comparison between ruling elite and entertainment-sports groups: Difference of ratio from the general population by Alphabetical Columns

Note: Vertical line (0) is the reference group (General population)

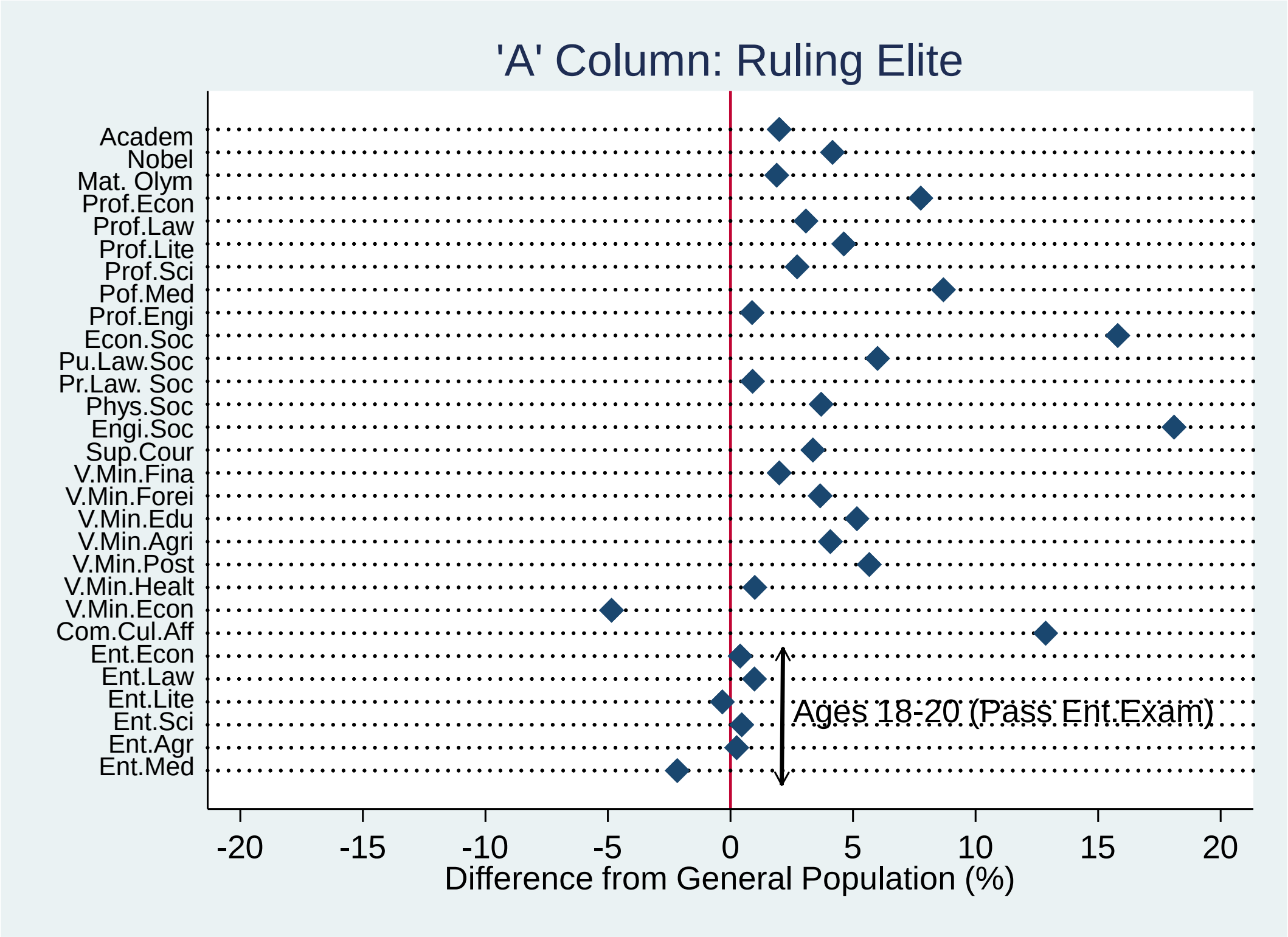

Fig4(a) Difference from the general population: Specific ruling elite groups.
Note: Vertical line (0) is the reference group (General population)

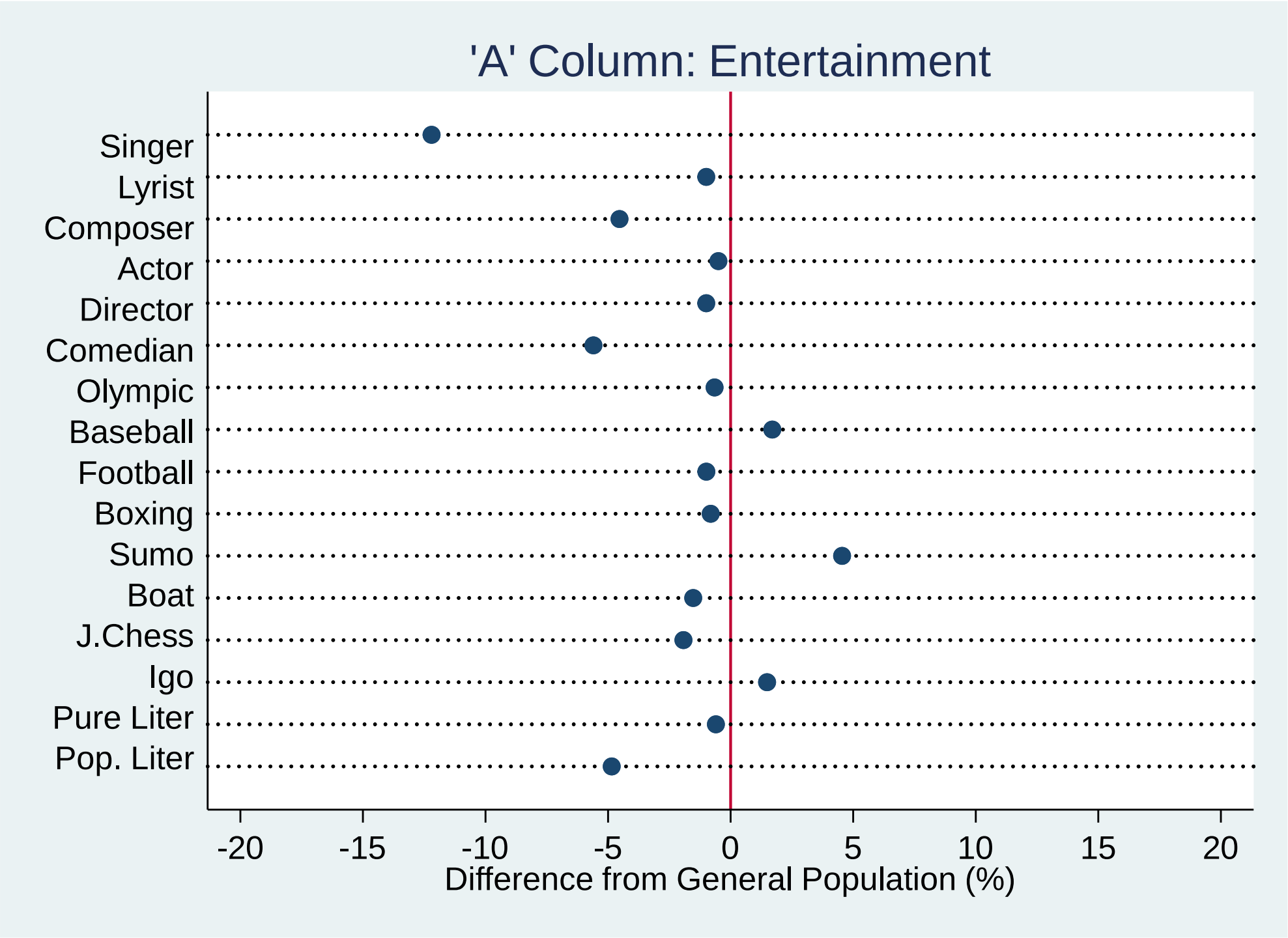

Fig4(b) Difference from the general population: Specific entertainment-sports groups.